\documentclass[a4paper,11pt]{article}
\usepackage{pos}

\title{B rare decays: theory overview}
\author*[a,b]{Marco Fedele}

\affiliation[a]{Institute for Theoretical Particle Physics, Karlsruhe Institute of Technology (KIT),\\
  Wolfgang-Gaede-Str. 1, D-76131 Karlsruhe, Germany}
\affiliation[b]{Departament de Física Teòrica, IFIC, Universitat de València – CSIC,
Parque Científico, Catedrático José Beltrán 2, E-46980 Paterna, Spain}

\emailAdd{marco.fedele@ific.uv.es}

\abstract{In this proceeding we will review the current theoretical status of rare $B$ decays. These decays are indeed excellent indirect probes for New Physics searches, and in the current situation where no new states have been directly observed at collider, they provide a fundamental and alternative approach in the quest for Physics beyond the Standard Model. We will focus on the following classes of decays: $B_q\to\tau\nu$, $B_q\to\mu\mu$, $B\to K^{(*)}\nu\bar\nu$, $B\to K^{(*)}\ell\ell$, $B_s\to\phi\ell\ell$ and $b\to s\gamma$. The most updated Standard Model predictions will be provided, highlighting which are the main sources of uncertainty, and what is the possibility for New Physics effects when confronting the theory numbers to current experimental results.}

\FullConference{Workshop Italiano sulla Fisica  ad Alta Intensità (WIFAI2023)\\
8-10 November 2023\\
Dipartimento di Architettura dell'Università Roma Tre, Rome, Italy\\}

\begin{document}
\maketitle

\section{Introduction}
Rare $B$ decays are excellent probes for New Physics (NP) searches. Given the current lack of direct production for NP states at present experimental facilities, alternative avenues must be explored to investigate potential extensions of the Standard Model (SM). One promising approach involves the meticulous examination of rare processes, wherein the presence of NP effects as intermediate, virtual states may become evident due to the already suppressed SM contribution.

Indeed, most of these rare $B$ decays are mediated by Flavour Changing Neutral Currents (FCNC), which are forbidden in the SM at tree-level. These processes occur at the loop-level and are therefore very rare, being generally both GIM- and CKM-suppressed. This proceeding aim to review the current theoretical status of the most promising rare $B$ decays. Precise experimental measurements are being confronted with accurate theoretical predictions in (and beyond) the SM, contributing to the ongoing search for NP effects.

\section{$B_q\to\tau\nu$}
The first class of decays here reviewed are the leptonic $B_q\to\tau\nu$ decays. While these processes are not mediated by an FCNC, they are nevertheless considered rare ones due to helicity suppression. Their Branching Ratio (BR) can be predicted in the SM in the following way:
%%%%%%%%%%%%%%%%
\begin{equation}
\mathcal{B}(B_q^+ \to \tau^+ \nu_\tau)^{\textrm {SM}} =\tau_{B_q^+} \frac{G_{F}^2 |V_{qb}|^{2} f_{B_q^+}^2 m_{B_q^+} m_\tau^2}{8\pi} \left( 1- \frac{m_\tau^{2}}{m_{B_q^+}^{2}} \right )^{2},  \qquad\qquad q = u,c
\end{equation}
%%%%%%%%%%%%%%%%
where $G_F$ is the Fermi constant, $\tau_{B_q^+}$ and $m_{B_q^+}$ denote the  $B_q^+$ meson lifetime and mass, respectively, and $m_\tau$ is the mass of the $\tau^+$ lepton. The main sources of theoretical uncertainty come from the CKM elements $|V_{qb}|$ and the $B_q^+$ meson decay constants $f_{B_q^+}$. Concerning the latter, the most precise measurements to date come from Lattice QCD (LQCD) and read $f_{B_c^+}=427(6)$~MeV and $f_{B^+}=190.0(1.3)$ MeV~\cite{FlavourLatticeAveragingGroupFLAG:2021npn}. Regarding the former, a long-standing discrepancy is currently present among inclusive and exclusive determinations of both $|V_{cb}|$ and $|V_{ub}|$~\cite{HFLAV:2022esi}; however, the unitarity of the CKM matrix allows for indirect extractions of these elements via global fits to all the other decays feeding in the so-called Unitarity Triangle Analysis (UTA). The latest predictions for these elements performed by the UTfit collaboration~\cite{UTfit:2022hsi} read $|V_{cb}|=42.22(51)\times 10^{-3}$ and $|V_{ub}|=3.70(11)\times 10^{-3}$. Using these input, the latest SM predictions for the two BRs read:
%%%%%%%%%%%%%%%%
\begin{equation}
\mathcal{B}(B_c^+\to \tau^+ \nu_\tau)^\mathrm{SM}=2.29(9)\times 10^{-2}\,,\qquad\qquad
\mathcal{B}(B^+\to \tau^+ \nu_\tau)^\mathrm{SM}=0.87(5)\times 10^{-4}\,.
\end{equation}
%%%%%%%%%%%%%%%% 

Going beyond the SM, this class of decays is sensitive to additional contribution from both vector and scalar currents, stemming from the NP operators
%%%%%%%%%%%%%%%%
\begin{equation}
O_{V_{L(R)}}= (\bar{q}_{L(R)}\gamma_\mu{b}_{L(R)})(\bar{\tau}_{L}\gamma_\mu \nu_L)\,,\qquad\qquad
O_{S_{L(R)}}= (\bar{q}_{R(L)}{b}_{L(R)})(\bar{\tau}_{R} \nu_L)\,.
\end{equation}
%%%%%%%%%%%%%%%%
We will denote with $C_i$ the respective couplings. These decays are excellent probes for NP effects coming from the scalar operators $O_{S_{L(R)}}$, due to the induced lift of the helicity suppression:
%%%%%%%%%%%%%%%%
\begin{align}
\mathcal{B}(B_q^+ \to \tau^+ \nu_\tau)= \mathcal{B}(B_q^+ \to \tau^+ \nu_\tau)^\mathrm{SM}\times\left|1-\left(C^q_{V_R} - C^q_{V_L}\right)+\left(C^q_{S_R} - C^q_{S_L}\right)\dfrac{m_{B_q}^2}{m_\tau(m_b+m_q)}\right|^2\,,
\end{align}
%%%%%%%%%%%%%%%%
Such contributions typically arise in models involving an additional Higgs doublet and/or a scalar Leptoquark (LQ) (for a study on current and future bounds see, e.g., Ref.~\cite{Zuo:2023dzn}).

\section{$B_q\to\mu\mu$}\label{sec:bqmumu}
Another class of helicity suppressed decays, which are moreover mediated by FCNC, consists of $B_q\to\mu\mu$ decays. The SM BR for these transitions is described by
%%%%%%%%%%%%%%%%
\begin{equation}
\mathcal{B}(B_q^0 \to \mu^+ \mu^-)^{\textrm {SM}} =\tau_{B_q^0} \frac{G_{F}^4 |V_{tb}^*V_{tq}|^{2} f_{B_q}^2 m_W^4 m_{B_q^0} m_\mu^2}{2\pi^5} \sqrt{1- \frac{4m_\mu^{2}}{m_{B_q^0}^{2}}} \left|C_{10}^{\rm q,SM}\right|^2, \qquad\quad q = d,s
\end{equation}
%%%%%%%%%%%%%%%%
where $C_{10}^{\rm q,SM}$ is the SM coupling associated to the axial current $Q_{10}^q = \frac{\alpha_{e}}{4\pi}(\bar{q}_L\gamma_{\mu}b_L)(\bar{\mu}\gamma^{\mu}\gamma_5\mu)$ and $m_W$ is the $W$ boson mass. Similarly to the previous case, also for this class of channels the main sources of uncertainty stem from the CKM elements $|V_{tq}|$ and the $B_q$ meson decay constants $f_{B_q}$, whose most precise determinations come from UTA analyses and LQCD, respectively. Concerning the CKM elements, the latest determinations read $|V_{td}|=8.59(11)\times 10^{-3}$ and $|V_{ts}|=41.28(46)\times 10^{-3}$~\cite{UTfit:2022hsi}; for the decay constants, we have $f_{B_d}=190.5(1.3)$~MeV and $f_{B_s}=230.1(1.2)$ MeV~\cite{FlavourLatticeAveragingGroupFLAG:2021npn}. Using these input, we can give the most precise and updated SM predictions for the two BRs:
%%%%%%%%%%%%%%%%
\begin{equation}
\mathcal{B}(B_d\to \mu^+ \mu^-)^\mathrm{SM}=9.48(36)\times 10^{-11}\,,\qquad\qquad
\mathcal{B}(B_s\to \mu^+ \mu^-)^\mathrm{SM}=3.47(14)\times 10^{-9}\,.
\end{equation}
%%%%%%%%%%%%%%%%

In the presence of NP effects, these class of decays is sensitive to additional contributions not only to the axial operator present in the SM, but also to the additional (pseudo)scalar operators,
%%%%%%%%%%%%%%%%
\begin{equation}
Q_{S}= \frac{\alpha_{e}}{4\pi}\frac{m_b}{m_W}(\bar{q}_{L}{b}_{R})(\bar{\ell}\ell)\,,\qquad\qquad
Q_{P}= \frac{\alpha_{e}}{4\pi}\frac{m_b}{m_W}(\bar{q}_{L}{b}_{R})(\bar{\ell}\gamma_5\ell)\,,
\end{equation}
%%%%%%%%%%%%%%%%
and to primed operators, obtained replacing $P_{L(R)}$ with $P_{R(L)}$. The modified expression for the BR, where the helicity suppression is again lifted in (pseudo)scalar contributions, reads
%%%%%%%%%%%%%%%%
\begin{align}
\mathcal{B}= \mathcal{B}^\mathrm{SM}\times \left( \left|
\frac{C_{10}^{\rm q,NP} - C_{10}^{\rm \prime q,NP}}{C_{10}^{\rm q,SM}}
+\dfrac{m_{B_q}^2}{2m_\mu m_b}\frac{C_{P}^{\rm q,NP} - C_{P}^{\rm \prime q,NP}}{C_{10}^{\rm q,SM}}\right|^2
+ \left| \sqrt{1-\frac{4m_\mu^2}{m_{B_q}^2}}\dfrac{m_{B_q}^2}{2m_\mu m_b}\frac{C_{S}^{\rm q,NP} - C_{S}^{\rm \prime q,NP}}{C_{10}^{\rm q,SM}}\right|^2
\right)\,.
\end{align}
%%%%%%%%%%%%%%%%
The current agreement between the SM predictions and the experimental measurements in these channels allows us to put stringent bounds on the involved NP couplings. We will comment on the most phenomenologically interesting ones in Sec.~\ref{sec:bsll}, in the context of global fits to $b\to s\ell\ell$ data.

\section{$B\to K^{(*)}\nu\bar\nu$}
In this section we will review the FCNC semi-leptonic decays $B\to K^{(*)}\nu\bar\nu$. They share the same kind of theoretical uncertainties presented for $B_s\to\mu\mu$ decays in Sec.~\ref{sec:bqmumu}, with additional ones stemming from the presence of form factors mediating the hadronic transitions present in these channels. Indeed, the hadronic matrix elements can be parameterized for the two channels as
%%%%%%%%%%%%%%%%
\begin{align}
\langle \bar{K}(k)|\bar{s}\gamma^\mu b|\bar{B}(p)\rangle &= \left[ (p+k)^\mu - \frac{m_B^2-m_K^2}{q^2}q^\mu \right] f_+(q^2) + \frac{m_B^2-m_K^2}{q^2}q^\mu f_0(q^2) \,,
\end{align}
%%%%%%%%%%%%%%%%
%%%%%%%%%%%%%%%%
\begin{align}
\langle \bar{K}^*(k)|\bar{s}\gamma^\mu (1-\gamma_5) b|\bar{B}(p)\rangle &= \epsilon_{\mu\nu\rho\sigma}\varepsilon^{*\nu}p^\rho k^\sigma{2V(q^2) \over m_B+m_{K^*}} 
 \nonumber \\
& -i \varepsilon_{\mu}^* (m_B+m_{K^*})A_1(q^2) + i (p+k)_\mu(\varepsilon^*q){A_2(q^2) \over m_B+m_{K^*}} \nonumber \\
& +i q_\mu(\varepsilon^*q){2m_{K^*} \over q^2}\left[ {m_B+m_{K^*} \over 2m_{K^*}} A_1(q^2) - {m_B-m_{K^*} \over 2m_{K^*}} A_2(q^2)-A_0(q^2)\right] \,.
\end{align}
%%%%%%%%%%%%%%%%
The form factors $f_0$ and $f_+$ involved in $B\to K$ transitions have been estimated via LQCD~\cite{Bailey:2015dka,Parrott:2022rgu} and combined in Refs.~\cite{Becirevic:2023aov,Bause:2023mfe}. Concerning the form factors $V$, $A_0$, $A_1$ and $A_2$ mediating $B\to K^*$ transitions, an estimate by means of Light-Cone Sum Rules (LCSR) has been in given in Ref.~\cite{Straub:2015ica}, which incorporates LQCD results from Ref.~\cite{Horgan:2013hoa}. We have now all the elements to write down the expression for the differential BRs, which read in the SM as~\cite{Buras:2014fpa}:
%%%%%%%%%%%%%%%%
\begin{align}
\frac{d\mathcal{B}}{dq^2}(B \to K\nu\bar\nu) =&\, \tau_B \frac{G_F^2\alpha_{\rm em}^2}{256\pi^5}\frac{\lambda_K^{3/2}}{m_B^3} |C_L^{\rm SM}|^2 |V_{tb}^*V_{ts}|^{2} [f_+(q^2)]^2\,, \\
\frac{d\mathcal{B}}{dq^2}(B \to K^*\nu\bar\nu) =&\, \tau_B \frac{G_F^2\alpha_{\rm em}^2}{128\pi^5}\frac{\lambda_{K^*}^{1/2}q^2}{m_B^3}(m_B+m_{K^*})^2 |C_L^{\rm SM}|^2 |V_{tb}^*V_{ts}|^{2} \bigg ( [A_1(q^2)]^2 \,, \\
& + \frac{32 m_{K^*}^2m_B^2}{q^2(m_B+m_{K^*})^2} [A_{12}(q^2)]^2 + \frac{\lambda_{K^*}}{(m_B+m_{K^*})^4} [V(q^2)]^2 \bigg)\,,
\end{align}
%%%%%%%%%%%%%%%%
where we have introduced the SM coupling $C_L^{\rm SM}=-6.32(7)$~\cite{Buras:2014fpa} defined as the universal, flavour-diagonal part of the coupling $C_L^{ij}$ mediating the operator $\mathcal{O}_L^{ij}=\frac{\alpha_{e}}{4\pi}(\bar{s}_L\gamma_{\mu}b_L)(\bar{\nu}_i\gamma^{\mu}(1-\gamma_5)\nu_j)$. Furthermore, $A_{12}$ is a linear combination of the form factors $A_1$ and $A_2$~\cite{Horgan:2013hoa}, and $\lambda_M\equiv \lambda(q^2, m_B^2, m_M^2)$ with $M=K,K^*$, is the K\"all\'en-function defined as $\lambda(a,b,c) = a^2 + b^2 + c^2 - 2(ab+ac+bc)$. We can give now the SM prediction for the integrated BRs of these decays~\cite{Bause:2023mfe,Allwicher:2023xba}:
%%%%%%%%%%%%%%%%
\begin{equation}
\mathcal{B}(B^\pm\to K^\pm \nu\bar\nu)=(4.44 \pm 0.30)\times 10^{-6}\,,\qquad
\mathcal{B}(B^\pm\to K^{*\pm} \nu\bar\nu)=(9.8 \pm 1.4)\times 10^{-6}\,.
\end{equation}
%%%%%%%%%%%%%%%%

Given the impossibility to flavour-tag the neutrinos at colliders, it is customary to express NP contributions to these channels as
%%%%%%%%%%%%
\begin{align}
R_{K^{(*)}}^{\nu\bar\nu} = \frac{{\mathcal B}(B\to K^{(*)} \nu\bar\nu)}{{\mathcal B}^{\rm SM}(B\to K^{(*)} \nu\bar\nu)} =&\ 1 + \sum_i
\frac{2 {\rm Re} [C_L^{\rm SM}(C_L^{ii} + C_R^{ii})]}{3|C_L^{\rm SM}|^2}  \,, \nonumber \\
& + \sum_{i,j} \left( \frac{|C_L^{ij} + C_R^{ij}|^2}{3|C_L^{\rm SM}|^2}
- \eta_{K^{(*)}} \frac{{\rm Re} [C_R^{ij}(C_L^{\rm SM}\delta^{ij} + C_L^{ij})]}{3|C_L^{\rm SM}|^2} \right)\,,
\end{align}
%%%%%%%%%%%%
where we have introduced $\eta_K=0$, $\eta_{K^*}=3.33(7)$, and $C_R^{ij}$ is the coupling of the operator $\mathcal{O}_R^{ij}=\frac{\alpha_{e}}{4\pi}(\bar{s}_R\gamma_{\mu}b_R)(\bar{\nu}_i\gamma^{\mu}(1-\gamma_5)\nu_j)$. The Belle II collaboration recently measured for the first time $R_{K}^{\nu\bar\nu} = 5.4 \pm 1.5$~\cite{Belle-II:2023esi}, obtaining a result $2.9\sigma$ larger than its SM prediction; when combined with previous upper limits, this result yields $R_{K}^{\nu\bar\nu} = 2.8 \pm 0.8$. Conversely, only upper limits relative to the $B^\pm\to K^{*\pm} \nu\bar\nu$ decay have been reported to date, with the best being set by the Belle collaboration at $R_{K^{*}}^{\nu\bar\nu} < 2.7 $ with 90\% C.L.~\cite{Grygier:2017tzo}. 

The interpretation of the potential excess observed by Belle II in the $K^\pm$ channel in terms of NP effects is not trivial, once confronted with current limits present in the $K^{*\pm}$ one~\cite{Bause:2023mfe,Allwicher:2023xba}. For instance, an explanation in terms of a flavour-universal NP contribution to $C_L$ would be clearly viable only after a decrease of the discrepancy in the $K^\pm$ channel. Introducing non-universal contributions is however strongly constrained by current data on ratios of muon to electron BRs in $b\to s$ transitions, see Sec.~\ref{sec:bsll}; a non-universal component would have to be therefore predominantly connected to $\nu_\tau$. Nevertheless, such components would yield, e.g. in a LQ scenario, to additional contributions in $b\to c\tau\nu$ transitions, hence needing to confront with data and anomalies in that sector as well~\cite{Buttazzo:2017ixm}.

\section{$B\to K^{(*)}\ell\ell$, $B_s\to\phi\ell\ell$}\label{sec:bsll}
The following class of decays comprises rare semi-leptonic $B$ decays involving charged leptons in the final states, namely $B\to K^{(*)}\ell\ell$ and $B_s\to\phi\ell\ell$, with $\ell=e,\mu$. Similarly to $B\to K^{(*)}\nu\bar\nu$, one of the main sources of uncertainty in these class of channels stem from the form factors, whose number grows to three in the case of $B \to K$ transitions and to seven in the cases of $B\to K^*$ and $B_s\to\phi$ ones, due to the inclusion of tensor matrix elements as well:
%%%%%%%%%%%%
\begin{align}
\langle\bar{K}(k)|\bar c\sigma_{\mu\nu} b|\bar B(p)\rangle =&\, i \left(p_\mu k_\nu - p_\nu k_\mu \right) \frac{2 f_T(q^2)}{m_B+m_K} \,, \\
\langle\bar{K}^*(k)|\bar c\sigma_{\mu\nu} b|\bar B(p)\rangle =&\, i \epsilon_{\mu\nu\alpha\beta} \left[ -\varepsilon^{*\alpha}(p+k)^\beta T_1(q^2) + \varepsilon^{*\alpha}q^\beta {m_B^2-m_{K^*}^2 \over q^2}[T_1(q^2)-T_2(q^2)] \right. \nonumber \\
& \left. \qquad + (\varepsilon^* q)p^\alpha k^\beta {2 \over q^2} \left( T_1(q^2)-T_2(q^2) - {q^2 \over m_B^2 - m_{K^*}^2} T_3(q^2) \right) \right] \,.
\end{align}
%%%%%%%%%%%%
Analogously to the previous class of decays, these additional form factors have been estimated in the LQCD for the former case~\cite{Bailey:2015dka,Parrott:2022rgu}, and in a combination of LQCD and LCSR for the latter ones~\cite{Horgan:2013hoa,Straub:2015ica}. 

However, a second source of uncertainty is introduced by non-local matrix elements involving the four-quark operator $Q^c_2 = (\bar{s}_L\gamma_{\mu} c_L)(\bar{c}_L\gamma^{\mu}b_L)$, particularly in proximity of the $c\bar c$ threshold, that yields non-factorizable power corrections from the time-ordered product
%%%%%%%%%%%%
\begin{align}
h_\lambda(q^2) = \frac{\epsilon^*_\mu(\lambda)}{m_B^2} \int d^4x\ e^{iqx} \langle \bar K^* \vert \mathcal{T}\{j^\mu_\mathrm{em}(x)
Q^{c}_{2} (0)\} \vert \bar B \rangle \,,
\end{align}
%%%%%%%%%%%%
with $j^\mu_\mathrm{em}(x)$ the electromagnetic (quark) current and $\lambda=\{0,+,-\}$ representing the helicity. While considerable progress has been made in estimating (at least part of) these amplitudes using light-cone sum rules \cite{Khodjamirian:2010vf,Khodjamirian:2012rm} and analyticity supplemented with perturbative QCD in the Euclidean $q^2$ region~\cite{Bobeth:2017vxj,Gubernari:2020eft,Gubernari:2022hxn}, calculating these hadronic contributions remains an open problem. Moreover, $h_\lambda(q^2)$ can mimic the presence of NP effects in these channels as it enters in the amplitudes as a Lepton Flavour Universal (LFU) shift to the coefficient $C_9$ mediating the operator $Q_{9} = \frac{\alpha_{e}}{4\pi}(\bar{s}_L\gamma_{\mu}b_L)(\bar{\mu}\gamma^{\mu}\mu)$~\cite{Ciuchini:2015qxb}, hence polluting the cleanness of the plethora of angular observables that can be defined for these three- and four-body decays beyond the usual BRs~\cite{Altmannshofer:2008dz,Descotes-Genon:2013vna}. To this end, Lepton Flavour Universality Violating (LFUV) ratios has been defined as $R_{K^{(*)}}=\mathcal{B}(B\to K^{(*)}\mu\mu)/\mathcal{B}(B\to K^{(*)}ee)$~\cite{Hiller:2003js}, where the uncertainties introduced by these non-local matrix elements largely cancel out yielding to the clean theoretical prediction of $R_{K^{(*)}}=1.00(1)$~\cite{Bordone:2016gaq}.

In the last decades an increasing number of so-called \emph{anomalies} has been measured by the LHCb collaboration, both in the BRs of $B\to K^{(*)}\mu\mu$ and $B_s\to\phi\mu\mu$, and in angular analyses of decays with vector mesons in the final state (see, e.g., Refs.~\cite{LHCb:2014cxe,LHCb:2020lmf,LHCb:2020gog,LHCb:2021zwz,LHCb:2021xxq} for the most updated results). While all these measurements are potentially plagued by the presence of non-hadronic matrix elements, a claim for NP effects coupled to the muon current was advocated due to observation by the LHCb collaboration of under-abundance of muon productions in the measurements of LFUV ratios $R_K$ and $R_{K^*}$~\cite{LHCb:2017avl,LHCb:2021trn}. However, after a recent re-analysis of LHCb data concerning these ratios~\cite{LHCb:2022vje}, there is no longer evidence for lepton-flavour violating NP effects in these channels~\cite{Ciuchini:2022wbq}. 

Nevertheless, the presence of LFU NP effects in this channel cannot be excluded yet, particularly in the context of global fits where all decays involving $b \to s \ell \ell$ transitions are taken into account. Indeed, remembering the constraints on $C_{10}$ coming from current data on $B_s \to \mu\mu$ mentioned in Sec.~\ref{sec:bqmumu}, the presence of LFU NP effects in $C_9$ is still not excluded, albeit with different significance according to the treatment of non-local hadronic uncertainties~\cite{Ciuchini:2022wbq,Greljo:2022jac,Alguero:2023jeh}.

\section{$b\to s\gamma$}
We conclude our review with the study of the radiative decays mediated by the $b\to s\gamma$ transition. Starting from the inclusive decay $B \to X_s \gamma$, the SM prediction for its BR is based on the equation
%%%%%%%%%%%%
\begin{align}
\mathcal{B}_{s\gamma}\equiv\mathcal{B}(B \to X_s \gamma)_{E_\gamma>E_0} = \mathcal{B}(B \to X_c \ell\nu) \left\vert \frac{V_{tb}^*V_{ts}}{V_{cb}}\right\vert^2 \frac{6e^2}{4\pi^2C} [P(E_0) + N(E_0)]\,,
\end{align}
%%%%%%%%%%%%
where $E_0=1.6$ GeV, $C$ is the so-called semi-leptonic phase-space factor, and $P(E_0)$ and $N(E_0)$ are the perturbative and non-perturbative contributions to the decay, respectively. The latest prediction for the former term is reaching the NNLO in QCD, combined with the most recent estimate for the former one~\cite{Gunawardana:2019gep}, yield to $\mathcal{B}_{s\gamma}=(3.40 \pm 0.17)\times 10^{-4}$~\cite{Misiak:2020vlo}, where the uncertainty stems from higher-order effects ($\pm3\%$), interpolation in $m_c$ ($\pm3\%$), and parametric non-perturbative effects ($\pm2.5\%$), which are added in quadrature. This result is in perfect agreement with current experimental measurements.

Of analogous interest are the associated exclusive decays $B_q \to V\gamma$, with $V$ being a vector meson like $K^*$ or $\phi$. For this class of decays it is possible to write the following observables:
%%%%%%%%%%%%
\begin{align}
\mathcal{B}(B_q \to V \gamma) &= \tau_{B_q} \frac{G_{F}^2 e^2 |V_{tb}^*V_{tq}|^{2} m_{B_q}^3 m_b^2}{128\pi^4} \left(1 - \frac{m_V^2}{m_B^2} \right)^3 (\left|C_{7}\right|^2 + \left|C_{7}'\right|^2)T_1(0) \,,\\
A_{\rm CP}(B_q(t) \to V \gamma) &= \frac{\Gamma(\bar B_q(t) \to \bar V \gamma) - \Gamma(B_q(t) \to V \gamma)}{\Gamma(\bar B_q(t) \to \bar V \gamma) + \Gamma(B_q(t) \to V \gamma)}\,,
\end{align}
%%%%%%%%%%%%%%%%
with $C_{7}$ being the coupling of the operator $Q_{7} = \frac{e}{16\pi^2}m_b\bar{s}_L\sigma_{\mu\nu}F^{\mu\nu}b_R$, and corresponding to the LO term of $P(E_0)$.

A combined analysis of both inclusive and exclusive radiative $B$ decays has been performed in Ref.~\cite{Paul:2016urs}, where the overall agreement of all the SM predictions with experimental measurements put very stringent constraints of potential NP effects in such channels.

\acknowledgments  This research was supported the BMBF grant 05H21VKKBA, from the Generalitat Valenciana (Grant PROMETEO/2021/071) and by MCIN/AEI/10.13039/501100011033 (Grant No. PID2020-114473GB-I00).

\bibliographystyle{JHEP}
\bibliography{hepbiblio}

\end{document}